\documentclass[a4paper,11pt]{article}

\usepackage{jheppub} 

\usepackage[T1]{fontenc} 

\title{\boldmath Non-extensive Statistical Mechanics and the Thermodynamic Stability of FRW Universe}

\author{Yang Liu}

\affiliation[a]{School of Physics and Astronomy, University of Nottingham, Nottingham NG7 2RD, UK}
\affiliation[b]{Nottingham Centre of Gravity, University of Nottingham, Nottingham NG7 2RD, UK}

\emailAdd{yang.liu@nottingham.ac.uk}

\abstract{In this article,  we  investigate the thermodynamic stability of the FRW  universe  for two examples, Tsallis entropy and loop quantum gravity, by considering non-extensive statistical mechanics. The heat capacity, free energy and pressure of the universe are obtained. For the Tsallis entropy model, we obtained the constraint for $\beta$, namely, $\frac{1}{2}<\beta<2$. The free energy of a thermal equilibrium universe must be less than zero. We suggest that the reason for the accelerated expansion of the universe is not due to Tsallis entropy. Similar results are obtained for loop quantum gravity. However, since the values of $\Lambda (\gamma)$ and $q$ cannot be determined in this model, the results become more subtle than that in the Tsallis entropy model. In addition, we compare the results for the universe with those for a Schwarzschild black hole.}

\begin{document} 
\maketitle
\flushbottom

\section{Introduction}
\label{sec:intro}
Although gravity is the most universal force in the universe and has been studied for more than a hundred years, black hole thermodynamics, which focuses on describing the entropy of black holes, gravitational phase transitions and the thermodynamic stability of black holes remains one of the most important research areas in modern gravity theory. The initial work on this subject was from several famous papers, which were based on standard thermodynamics and Boltzmann-Gibbs statistical mechanics [1,2,3,4]. Following on from these famous papers, researchers generalized and modified the area law of entropy [5,6,7,8,9,10]. \\ 
One modification of the area law of entropy came from Gibbs's arguments that the Boltzmann-Gibbs (BG) theory cannot be applied in systems with divergency in the partition function, such as the gravitational system. Therefore,  thermodynamic entropy of such non-standard systems cannot be described by an extensive entropy but must instead be generalized to a  non-extensive entropy.  Considering this conclusion and using statistical arguments, Tsallis and Cirto have pointed out that the mathematical expression for  thermodynamic entropy of a black hole can be modified as [6]
\begin{equation}\label{eq:1.1}
S_h = \alpha A^\beta,
\end{equation}
where $A$ is the black hole horizon area, $\alpha$ an unknown constant and $\beta$ the Tsallis parameter or “non-extensive” parameter, which is a real parameter that quantifies the degree of nonextensivity [6,11].\footnote{The terminology is subtle. We should point out that even if the theory is called “nonextensive statistical mechanics”, Tsallis entropy is extensive (but it is non-additive).} If one takes $\beta=1$ and $\alpha= \frac{1}{4 L^2_p} $, then the area law of entropy can be restored. Sheykhi then generalized this result to FRW universe and obtained the modified Friedmann equations [11].\\
Loop quantum gravity (LQG) or quantum geometry is one of the candidates used in quantum gravity theory. This theory can provide us with an estimate of the microstates of a black hole with a given classical area (A) and at the kinematical level has led to a precise computation of its entropy [7]. In ref.[7,9], researchers have explored the basic results for the entropy of black holes from non-extensive statistical mechanics in LQG. In ref.[9], in particular, the authors have examined the thermodynamic stability of black holes and the basic method derived there is used in this article. Combining the results of non-extensive statistics and quantum geometry, the authors have suggested an alternative mathematical expression for black hole entropy:
\begin{equation}\label{eq:1.2}
S_q = \frac{k_B}{1-q} [ \exp \left( (1-q)\Lambda (\gamma) \frac{A}{4 L^2_p} \right) -1 ],
\end{equation}
where $\gamma$ is Barbero-Immirzi parameter and  $\Lambda (\gamma) = \frac{ln 2}{\gamma \pi \sqrt{3}} $ [9]. In classical theory the parameter $\gamma$ represents a one-parameter family of canonical transformation of the canonical variables, i.e., for every value of $\gamma$ the classical equations of motion of general relativity are valid [7]. While $\gamma$ represents a quantization ambiguity of the theory in the quantum theory [7].\\
However, all of these studies [5,6,7,8,9,10] focus on the thermodynamic properties of black holes. In particular, in ref.[5,8,10], the authors focused on black hole thermodynamics with Rényi entropy, which is another example of nonextensive statistical mechanics and we will not consider in this article. In this article, we use the methods in ref.[6,7,9] to investigate the thermodynamic stability of the FRW universe. We explore two cases by way of example: Tsallis entropy and loop quantum gravity which we have discussed briefly in previous paragraphs. In section 2, we introduce the basic results of using non-extensive statistics for Tsallis entropy and loop quantum gravity. In section 3 and section 4, we investigate the thermodynamic stability of the FRW universe for these two cases. In section 5, we present the results and compare them with those for black holes.\\ 

\section{Non-extensive Statistical Mechanics and the Modified Area Law of Entropy}
In this section, we will briefly review the basics of non-extensive statistical mechanics and the modified area law of entropy.
\subsection{non-extensive statistical mechanics and q-entropy}
First of all, at the level of statistical mechanics, including the influence of deviations on the probability of a given microstate of the potential quantum mechanical system, is the idea of introducing the concept of $q$-entropy [7,9]. The parameter $q$ is the entropic index. We have $0 <p_i<1$, and thus $p^q_i>p_i$ for $q <1$ and $p^q_i<p_i$ for $q >1$. This means that for $q>1$, frequent events with probability close to 1 are relatively enhanced, while for $q<1$ rare events with a probability close to zero are relatively enhanced. Considering this fact, Tsallis suggested the following definition for $q$-entropy:
\begin{equation}\label{eq:2.1}
S_q = \frac{1- \sum_{i=1}^{W} p^q_i}{q-1}, 
\end{equation}
where $p_i$ is the probability of the system existing in a given microstate, $W$ the total number of available microstates and the $q$ the Tsallis parameter. The parameter $q$ is a real number and when $q$ approaches 1, the Tsallis entropy will become the Shannon entropy:
\begin{equation}\label{eq:2.2}
S = -\sum_{i=1}^{\Omega} p_i \, ln p_i,  
\end{equation}
where $p_i$ is the probability of the $i$-th microstate and $\Omega$ is the total number of microstates of the system under consideration [7]. The related branch of statistical mechanics is known as non-extensive statistical mechanics (NESM) [7,9]. For equal probability $p_i = \frac{1}{\Omega}$ for all $i$ and eq.$(2.1)$ reduces to [7,9]
\begin{equation}\label{eq:2.3}
 S_q = ln_q \Omega,  
\end{equation} 
where $ln_q x =\frac{1- x^{1-q}}{q-1} $ is known as the $q$-logarithm. If we consider a group of particles whose spin sequence is $(j_1,..., j_N)$, due to basic quantum mechanics, the setting $\Omega$ in eq.$(2.3)$ can be given by $\Omega(j_1,..., j_N) = \prod_{l=1}^{N} (2 j_l + 1)$ [7]. Furthermore, if we consider $j_1=...=j_N=s$, then eq.$(2.3)$ becomes [7]
\begin{equation}\label{eq:2.4}
S_q = \frac{(1+2s)^{(1-q)N}-1}{1-q}. 
\end{equation} 

\subsection{Tsallis and Cirto's modified area law of entropy}
One modification of the area law of entropy comes from the Gibbs´s arguments. He pointed out that if the partition function of a system is divergent, as it is for a  gravitational system,  then  the Boltzmann Gibbs (BG) theory cannot be applied. The thermodynamic entropy of such non-standard systems must be described instead by non-extensive entropy rather than extensive entropy [12,13,14,15]. In view of this fact and considering the statistical argument, Tsallis and Cirto have suggested that the thermodynamic entropy of a black hole should be modified as
\begin{equation}\label{eq:2.5}
S_h = \alpha A^\beta,
\end{equation}
where $A$ is the black hole horizon area, $\alpha$ an unknown constant and $\beta$ the Tsallis parameter or “non-extensive” parameter, which is a real parameter that quantifies the degree of nonextensivity [5]. In this paper, we will call eq.$(2.5)$ the "Tsallis entropy". If one takes $\beta=1$ and $\alpha= \frac{1}{4 L^2_p} $, then the area law of entropy can be restored. In fact, under this limit, the power-law distribution of probability becomes irrevelant, and the system can be simply described by the ordinary distribution of probability [5,11]. Sheykhi's generalisation of this result to the FRW universe then leads to the modified Friedmann equations [11]. \\

\subsection{q-entropy for a black hole in loop quantum gravity}
Let us briefly discuss the basic structures of loop quantum gravity (LQG) of black holes [7,9]. The quantum geometry of the black hole horizon cross-section in the LQG is described topologically by two spheres with defects (usually called punctures), which have their “spin” quantum number determined by the edge of the spin network, representing the bulk quantum geometry [7,9]. The quantum area of such a black hole with spin quantum numbers $j_1,...,j_N$ for the $N$ punctures is given by [7,9]
\begin{equation}\label{eq:2.6}
A_{qu} = 8 \pi \gamma L^2_p \sum^{N}_{l=1} \sqrt{j_l (j_l + 1)},
\end{equation}
where $L_p$ is the Planck length and the number of microstates is
\begin{equation}\label{eq:2.7}
\Omega(j_1,..., j_N) = \prod_{l=1}^{N} (2 j_l + 1).
\end{equation}
Therefore, we actually have an a system with $N$ statistically independent spin quantum $j_1,...,j_N$. Considering all possible spin permutations, the total number of microstates for a black hole with a classical area $A$ is obtained, and thus $A_{qu}  = A \pm O(L^2_p)$.\\
In black hole thermodynamics, the minimal quantum of area that is permissible is one with spin $\frac{1}{2}$. In view of this, it is not hard to derive that the number of possible states is described by the case $ j_l = \frac{1}{2} \, \forall l \in [1, s] $ [9]. When $N \rightarrow \infty$, statistical mechanics can be applied to the system. Therefore, if we put $n_\frac{1}{2}= N, n_\frac{2}{2}= ...= n_\frac{s}{2} =0 $, then the $q$-entropy of a black hole as stated in eq.$(2.4)$ is obtained:
\begin{equation}\label{eq:2.8}
S_q = \frac{2^{(1-q)N}-1}{1-q},
\end{equation}
as demonstrated in ref.[7,9,16]. In addition, at this stage, $ j_l = \frac{1}{2} \, \forall l \in [1, s] $ and the area is described by [9]
\begin{equation}\label{eq:2.9}
A = 4 \pi \gamma L^2_p N \sqrt{3}.
\end{equation} 
In view of eqs.$(2.8)$ and $(2.9)$, then the entropy of the black hole can be written as
\begin{equation}\label{eq:2.10}
S_q = \frac{1}{1-q} [ \exp \left( (1-q)\Lambda (\gamma) \frac{A}{4 L^2_p} \right) -1 ],
\end{equation} 
where we have set $k_B =1$ and $\Lambda (\gamma) = \frac{ln 2}{\gamma \pi \sqrt{3}} $. In the $q \rightarrow 1$ limit, the standard Bekenstein-Hawking entropy is reproduced by both the Tsallis formulas and by $\gamma = \frac{ln 2}{\pi \sqrt{3}} $ as [9]:
\begin{equation}\label{eq:2.11}
S_{BH} = \frac{A}{4 L^2_p }.
\end{equation}

\section{Thermodynamic Stability of the FRW Universe from Tsallis Entropy}
We consider the Friedmann-Robertson-Walker (FRW) metric
\begin{equation}\label{eq:3.1}
ds^2 = h_{\mu \nu} dx^{\mu} dx^{\nu} + \tilde{r}^2 (d \theta^2 + sin^2 \theta d\phi^2),
\end{equation}
where $\tilde{r}=a(t)r$, $x_0=t$, $x_1=r$, and $h_{\mu \nu}=diag (-1,\frac{a^2}{1-k r^2} )$ represents the two dimensional metric. The situations for $k =0$, $1$, $-1$ corresponds to open, flat, and closed universes, respectively. We also assume that the physical boundary of the universe consistent with the laws of thermodynamics is the apparent horizon with radius [11]
\begin{equation}\label{eq:3.2}
\tilde{r}_A = \frac{1}{\sqrt{H^2 + \frac{k}{a^2} }}.
\end{equation}
The temperature associated with the apparent horizon can be defined as [17] 
\begin{equation}\label{eq:3.3}
T_h = \frac{\kappa}{2 \pi} = -\frac{1}{2 \pi \tilde{r}_A} (1- \frac{\dot{\tilde{r}}_A}{2 H \tilde{r}_A}),
\end{equation}
where $\kappa$ is the surface gravity. If $\dot{\tilde{r}}_A \ll 2 H \tilde{r}_A$ and in order to avoid negative temperature, we can define the temperature as [18]
\begin{equation}\label{eq:3.4}
T_h = \frac{1}{2 \pi \tilde{r}_A}.
\end{equation}
In this paper, we assume that the local equilibrium holds. Therefore, the thermal system bounded by the apparent horizon remains in equilibrium and the temperature of the system $T_m$ must be uniform and equal to the temperature of its boundary $T_h$, $T_m = T_h$ [11,19]. In the following section, we will not differentiate between these two temperatures.\\ 
For brevity, we assume the total energy content of the universe inside a $3$-dimensional sphere of radius $\tilde{r}_A$, is $E=\rho V$ where $\rho$ is the average density of the universe and $V=\frac{4 \pi}{3} \tilde{r}^3_A $ is the volume enclosed by a $3$-dimensional sphere having the area of apparent horizon $A =4 \pi \tilde{r}^2_A $ [11]. Then, based on thermodynamics, the heat capacity can be obtained as:
\begin{equation}\label{eq:3.5}
\begin{aligned}
C_U= T_h \frac{\partial S_h}{\partial T_h }=
& \frac{1}{2 \pi \tilde{r}_A} \frac{\partial (\alpha A^\beta)}{\partial (\frac{1}{2 \pi \tilde{r}_A})}=- 2\beta \alpha (4 \pi)^\beta \tilde{r}^{2\beta}_A.\\
\end{aligned}
\end{equation}
If we  regard the universe purely as a gravitational system, i.e., $C_U<0$, then we get $\beta > 0$. To investigate the global thermodynamical stability of the universe using Tsallis entropy, we need to consider its free energy $F_{U}$. $F_{U}$ can be defined as
\begin{equation}\label{eq:3.6}
F_{U} = E - T_h \times S_h = \frac{4 \pi}{3} \rho \tilde{r}^3_A - \frac{\alpha}{2 \pi} (4 \pi)^\beta (2\beta -1) \tilde{r}^{2\beta -1}_A.
\end{equation}  
In order to obtain the extremum of $F_{U}$, we set $\frac{\partial F_U}{\partial \tilde{r}_A} = 0$. Then we obtain
\begin{equation}\label{eq:3.7}
\rho = \frac{\alpha}{2 \pi} (4 \pi)^{\beta-1} (2\beta -1) \tilde{r}^{2\beta -4}_{A_0},
\end{equation}
where $\tilde{r}_{A_0}$ is the stationary point. Furthermore, we obtain $\frac{\partial^2 F_{U}}{\partial \tilde{r}^2_A}$ at $\tilde{r}_A = \tilde{r}_{A_0}$:
\begin{equation}\label{eq:3.8}
\frac{\partial^2 F_{U}}{\partial \tilde{r}^2_A} = 8 \pi \rho \tilde{r}_{A_0} - \frac{\alpha}{2 \pi} (4 \pi)^{\beta} (2\beta -1) (2\beta -2)\tilde{r}^{2\beta -3}_{A_0},  
\end{equation}
i.e.
\begin{equation}\label{eq:3.9}
\frac{\partial^2 F_{U}}{\partial \tilde{r}^2_A} = \frac{\alpha}{2 \pi} (4 \pi)^{\beta} (2\beta -1) (4- 2\beta)\tilde{r}^{2\beta -3}_{A_0}.  
\end{equation}
If we require that the universe is thermodynamically stable, so that $F_{U}$ must have minimum, then $\frac{\partial^2 F_{U}}{\partial \tilde{r}^2_A}$ must be larger than 0. Therefore, we have
\begin{equation}\label{eq:3.10}
 (2\beta -1) (4- 2\beta) > 0. 
\end{equation}
The only possibility is $\frac{1}{2}<\beta<2$. The area law $S=\frac{A}{4 L^2_p}  $ meets this requirement.\\
At $\tilde{r}_A = \tilde{r}_{A_0}$, the minimum of free energy $F_{U,m}$ is given by
\begin{equation}\label{eq:3.11}
F_{U,m} = \frac{4 \pi}{3} \rho \tilde{r}^3_{A_0} - \frac{\alpha}{2 \pi} (4 \pi)^\beta (2\beta -1) \tilde{r}^{2\beta -1}_{A_0} = - \frac{4 \alpha}{3}  (2\beta -1) (4 \pi)^{\beta -1} \tilde{r}^{2\beta -1}_{A_0}.
\end{equation}
Since $\frac{1}{2}<\beta<2$, we can conclude that $F_{U,m}$ must be smaller than zero. Throughout most of the history of the Universe (in particular the early Universe), the reaction rates of particles in the thermal bath, $\Gamma_{int}$, were much greater than the expansion rate, $H$, and local thermal equilibrium (LTE) should have been maintained [20]. Therefore, $F_{U,m}$ is the value of free energy of the Universe.\\
Furthermore, based on ref.[21], we have 
\begin{equation}\label{eq:3.12}
F=-k_B T ln_q Z_q, 
\end{equation}
and
\begin{equation}\label{eq:3.13}
p=k_B T \frac{\partial}{\partial V} ln_q Z_q, 
\end{equation}
where $Z_q$ is the partition function in non-extensive statistical mechanics, whose value does not need to be known for the purposes of this paper. Therefore we have
\begin{equation}\label{eq:3.14}
p=- T_h \frac{\partial}{\partial V} (\frac{F_U}{T_h}).
\end{equation}
Considering $V=\frac{4 \pi}{3} \tilde{r}^3_A$ and using eq.$(3.4)$, we obtain that
\begin{equation}\label{eq:3.15}
p_U=- \frac{1}{4\pi \tilde{r}^3_A } \frac{\partial}{\partial \tilde{r}_A} (\tilde{r}_A F_U) = - \frac{4 \rho}{3} + \frac{\alpha \beta}{\pi} (2\beta -1) (4\pi)^{\beta -1}  \tilde{r}^{2\beta -4}_A.
\end{equation}
At $\tilde{r}_A = \tilde{r}_{A_0}$ and in view of eq.$(3.7)$, then the pressure of the universe is then:
\begin{equation}\label{eq:3.16}
p_0= \frac{\alpha}{\pi} (\beta - \frac{2}{3}) (2\beta -1) (4\pi)^{\beta -1}  \tilde{r}^{2\beta -4}_{A_0}
\end{equation}
If we want to explain the accelerated expansion of the universe, then $p < 0$, i.e. $\frac{1}{2}<\beta<\frac{2}{3}$. However, this result is incompatible with the area law of entropy. If BH entropy indeed holds, i.e., $\beta =1$, then we suggest that the reason for the accelerated expansion of the universe is not due to Tsallis entropy.\\
Moreover, we can consider the thermodynamic stability of a Schwarzschild black hole. For a Schwarzschild black hole, we have [9]
\begin{equation}\label{eq:3.17}
A = 16 \pi M^2 = 4 \pi R^2_s,
\end{equation} 
since, $M=\frac{1}{2} R_s $. Here, $R_s$ is the radius of the horizon. Based on thermodynamics, 
\begin{equation}\label{eq:3.18}
\frac{1}{T_{BH}} = \frac{\partial S_h}{\partial E} = \frac{\partial S_h}{\partial M}.
\end{equation} 
Considering eqs.$(1.1)$ and $(3.12)$, we have
\begin{equation}\label{eq:3.19}
T_{BH} = \frac{1}{4 \alpha \beta (4 \pi)^\beta} R^{1-2\beta}_s.
\end{equation}
Then we can obtain the heat capacity of the black hole:
\begin{equation}\label{eq:3.20}
C_{BH}= T_{BH} \frac{\partial S_h}{\partial T_{BH} }=  \frac{2 \alpha \beta (4 \pi)^\beta}{1- 2\beta } R^{2\beta}_s.
\end{equation}
Since a Schwarzschild black hole is a purely gravitational system, $C_{BH}$ must be less than zero, so that $\frac{ \beta}{1- 2\beta }$ is smaller than zero. The two possibilities are either $\beta > \frac{1}{2}$ or $\beta < 0$. Again, if the area law holds, i.e., $ \beta = 1 $, then the former is the only valid possibility, which is consistent with the result for the Universe, i.e., $\frac{1}{2} < \beta < \frac{2}{3}$.\\
Furthermore, the free energy of a black hole $F_{BH}$ can also be obtained as:
\begin{equation}\label{eq:3.21}
F_{BH}=  M-T_{BH} \times S_h = (\frac{1}{2} - \frac{1}{4\beta}) R_s = (1- \frac{1}{2\beta}) M,
\end{equation}
which defines that the entropy of a black hole increases monotonically with both the radius of horizon and the mass of the black hole. It is evident that the free energy of a black hole according to this theory must be larger than zero.

\section{Thermodynamic Stability of the FRW Universe from Loop Quantum Gravity}
In this section, we will consider the thermodynamic stability of the FRW universe from the point of view of loop quantum gravity. The method is similar to that in section 3.\\
In view of eq.$(2.10)$ and $A =4 \pi \tilde{r}^2_A$, we have
\begin{equation}\label{eq:4.1}
S_q= \frac{1}{1-q} [ \exp \left( (1-q)\Lambda (\gamma) \frac{\pi \tilde{r}^2_A }{ L^2_p} \right) -1 ]. 
\end{equation}
Then we have
\begin{equation}\label{eq:4.2}
\frac{\partial S_q}{\partial \tilde{r}_A} = 2 \frac{\Lambda (\gamma) \pi \tilde{r}_A}{L^2_p} \exp \left( (1-q)\Lambda (\gamma) \frac{\pi \tilde{r}^2_A }{ L^2_p} \right) 
\end{equation}
and
\begin{equation}\label{eq:4.3}
\frac{\partial \tilde{r}_A}{\partial T_h} =\frac{\partial \tilde{r}_A}{\partial \frac{1}{2\pi \tilde{r}_A} } = -2\pi \tilde{r}^2_A.
\end{equation}
From this the heat capacity of the universe can be obtained as:
\begin{equation}\label{eq:4.4}
C_u= T_h \frac{\partial S_q}{\partial T_h } =T_h \frac{\partial S_q}{\partial \tilde{r}_A} \frac{\partial \tilde{r}_A}{\partial T_h} =- \frac{2 \Lambda (\gamma) \pi \tilde{r}^2_A}{L^2_p} \exp \left( (1-q)\Lambda (\gamma) \frac{\pi \tilde{r}^2_A }{ L^2_p} \right).
\end{equation}
Therefore, the heat capacity $C_u$ derived from loop quantum gravity must be less than zero.\\
Based on thermodynamic considerations,another expression of $T_h$ can be obtained as:
\begin{equation}\label{eq:4.5}
\frac{1}{T_h} = \frac{\partial S_q}{\partial M } = \frac{1}{2 \rho L^2_p \tilde{r}_A } \exp \left( (1-q)\Lambda (\gamma) \frac{\pi \tilde{r}^2_A }{ L^2_p} \right).
\end{equation}
Using eq.$(4.5)$, we can also derive the free energy of the universe $F_u$:
\begin{equation}\label{eq:4.6}
F_u = E - T_h \times S_q = \frac{4 \pi \rho}{3} \tilde{r}^3_A - \frac{2 \rho L^2_p \tilde{r}_A }{1-q} [1 - \exp \left( -(1-q)\Lambda (\gamma) \frac{\pi \tilde{r}^2_A }{ L^2_p}  \right)].
\end{equation}
In order to obtain the extremum of $F_u$, we set $\frac{\partial F_u}{\partial \tilde{r}_A} = 0$. Then we obtain
\begin{equation}\label{eq:4.7}
\begin{aligned}
\frac{\partial F_u}{\partial \tilde{r}_A} & = 4\pi \tilde{r}^2_A \rho  [1- \Lambda (\gamma) \exp \left( -(1-q)\Lambda (\gamma) \frac{\pi \tilde{r}^2_A }{ L^2_p}  \right)] \\
&-\frac{2}{1-q} \tilde{r}^2_A \rho [1- \exp \left( -(1-q)\Lambda (\gamma) \frac{\pi \tilde{r}^2_A }{ L^2_p}  \right)] =0.
\end{aligned}
\end{equation}
$F_u$ has two stationary points. The first one is $\tilde{r}_{A_1} = 0$. The second one satisfies the following relation
\begin{equation}\label{eq:4.8}
\tilde{r}^2_{A_2} = \frac{L^2_p}{(1-q) \pi \Lambda (\gamma) } ln \left( \frac{4 \pi \Lambda (\gamma) - 4 \pi \Lambda (\gamma) q-2 }{ 4 \pi - 4 \pi q-2 } \right).
\end{equation}
For $\tilde{r}_{A_1} = 0$, we have $F_{u_1}=0$. While for $\tilde{r}_{A_2}$, we have
\begin{equation}\label{eq:4.9}
F_{u_2} = 4 \pi \rho L^2_p [\frac{1}{3(1-q) \pi \Lambda (\gamma) } ln \left( \frac{4 \pi \Lambda (\gamma) - 4 \pi \Lambda (\gamma) q-2 }{ 4 \pi - 4 \pi q-2 } \right) - \frac{\Lambda (\gamma) -1 }{2 \pi \Lambda (\gamma) - 2 \pi \Lambda (\gamma) q-1}].
\end{equation}
When $q \rightarrow 1$, $\Lambda (\gamma) \rightarrow 1 $, we have $F_{u_2} \rightarrow 0$. The minimum value of free energy $F_u$ depends on the values of $\frac{1}{3(1-q) \pi \Lambda (\gamma) } ln \left( \frac{4 \pi \Lambda (\gamma) - 4 \pi \Lambda (\gamma) q-2 }{ 4 \pi - 4 \pi q-2 } \right)$ and $\frac{\Lambda (\gamma) -1 }{2 \pi \Lambda (\gamma) - 2 \pi \Lambda (\gamma) q-1}$. However, these values cannot be determined since $\Lambda (\gamma)$ and $q$ are not determined in this theory.\\
In a similar way, we can also obtain the pressure of the universe in loop quantum gravity. After some rearrangement of terms, we have
\begin{equation}\label{eq:4.10}
\begin{aligned}
p= &- \frac{\rho}{3} +  \frac{\rho L^2_p}{1-q} \frac{1}{2 \pi \tilde{r}^2_A } [1- \exp \left( -(1-q)\Lambda (\gamma) \frac{\pi \tilde{r}^2_A }{ L^2_p}  \right) ] \\
& - \rho [1- \Lambda (\gamma) \exp \left( -(1-q)\Lambda (\gamma) \frac{\pi \tilde{r}^2_A }{ L^2_p}  \right) ]\\
& + \frac{\rho}{2 \pi (1-q)} [1- \exp \left( -(1-q)\Lambda (\gamma) \frac{\pi \tilde{r}^2_A }{ L^2_p}  \right) ].
\end{aligned}
\end{equation}
Two possibile solutions exist. For $\tilde{r}_{A_1} = 0$, we have
\begin{equation}\label{eq:4.11}
p_1 = - \frac{4\rho}{3} [1- \frac{1}{2}\Lambda (\gamma) ].
\end{equation}
If $p_1$ is the pressure of the universe and we want to explain the accelerated expansion of the universe, i.e. $p < 0$, so that $\Lambda (\gamma)<2$. For $\tilde{r}_{A_2}$, we can obtain the other possibility $p_2$. We will not list the expression for $p_2$ since it is too lengthy. Which one of these two possibilites is the real pressure depends on the values of the other two parameters, namely, $\Lambda (\gamma)$ and $q$.\\
We should mention that in ref.[9], the authors have discussed the thermodynamic stability of a Schwarzschild black hole. The expressions for $T_{bh}$, $C_{bh}$ and $F_{bh}$ of a black hole have been obtained. Based on ref.[9], the following results are discussed. \\
Firstly, contrary to the asymptotically case at $(q \rightarrow 1)$, the Schwarzschild-Tsallis $(q>1)$ black hole temperature no longer decreases monotonically with mass. Instead it reaches a minimum at $m =m_{min}$; $T=T^{min}_H$. For $T_H<T^{min}_H$, black holes cannot exist and the space is filled with pure radiation, where $T^{min}_H = T_p \sqrt{\frac{(q-1)e}{8 \pi \Lambda (\gamma) }}$ and $m_{min} =[8 \pi (q-1) \Lambda (\gamma) ]^{\frac{-1}{2}} $.\\
Secondly, when $0 <m <m_{min}$ and $q >1$ or $(m >0$ and $q \le 1)$, the specific heat of system is negative, while for $m > m_{min}$ and $q >1$, it is positive. In this case $q >1$, the temperature has the a minimum at $T_H=T^{min}_H$.\\
Thirdly, the larger the black hole, the more thermodynamically stable it is. To ensure a positive heat capacity $C_{bh} \ge 0$ as well as physically valid further thermodynamic quantities, the inequalities $q >1$ and $m > m_{min}$ must be satisfied and $C_{bh} < 0 $ must hold for the smaller black holes with $ m <m_{min}$; these situations are thermodynamically unstable since they are not representative of thermal equilibrium with a bath of thermal radiation.\\
From the above we can conclude that, in a loop quantum gravity model, the thermodynamic stability of a Schwarzschild black hole is determined by whether the value of $(1-q)$ is larger than or smaller than zero. The precise results cannot be determined easily for the universe since they are dependent on the concrete values of $\Lambda (\gamma)$ and $q$, which cannot been derived in this theory.

\section{Conclusions and Outlook}
One of the most important research areas in modern gravity theory is black hole thermodynamics, which focuses on describing the entropy of black holes, gravitational phase transitions and the thermodynamic stability of black holes. The initial work on these issues started from several famous papers, which were based on standard thermodynamics and Boltzmann-Gibbs statistical mechanics [1,2,3,4].\\
A significant modification for the area law of entropy came from the Gibbs himself who emphasised  that the Boltzmann-Gibbs (BG) theory cannot be applied in systems with divergence in the partition function such as in the gravitational system. Therefore, thermodynamic entropy of such non-standard systems cannot be described by an extensive entropy but must instead be described using the more generalized concept of non-extensive entropy. All previous studies on this subject focus on the thermodynamic properties of black holes [5,6,7,8,9,10]. In ref.[6], Tsallis and Cirto suggested that the mathematical expression for thermodynamic entropy of a black hole should be modified as $S_h = \alpha A^{\beta}$, where $A$ is the black hole horizon area, $\alpha$ an unknown constant and $\beta$ the Tsallis parameter [6]. Futhermore, the authors of ref.[7,9] explored the basic results for the entropy of black holes from non-extensive statistical mechanics in Loop Quantum Gravity. In addition, ref.[5,8,10] concentrated on black hole thermodynamics with Rényi entropy, which is another important example of nonextensive statistical mechanics. However, in this article, we have investigated the basic thermodynamic stability of the FRW universe using two specific examples: Tsallis entropy and loop quantum gravity.\\ 
For the model using Tsallis entropy, in which we regard the universe as a purely gravitational system, the heat capacity of the universe is derived as being smaller than zero which means the universe is not stable. If we require that the universe is thermodynamically stable so that the free energy of the universe $F_U$ must have minimum, then we can derive the constraint for parameter $\beta$, namely, $\frac{1}{2}<\beta<2$. Furthermore, throughout most of the history of the Universe, local thermal equilibrium can be maintained [20], then the minimum of free energy $F_{U,m}$ must be less than zero. The pressure of the universe has also been obtained. In order to explain the accelerated expansion of the universe, i.e. $p<0$, then we have $\frac{1}{2}<\beta<\frac{2}{3}$. However, this result is inconsistent with the area law of entropy. Therefore, we suggest that the reason for accelerated expansion of the universe is not due to Tsallis entropy.  For comparision with the above conclusions for the universe, we also obtained the basic thermodynamic properties of a Schwarzschild black hole. Since a Schwarzschild black hole is a purely gravitational system, $C_{BH} $ must be less than zero, so that $\beta >\frac{1}{2} $, which is consistent with the result of the universe. The entropy of a black hole increases monotonically with both the radius of its event horizon and its mass, which are both greater  than zero. \\
Similar methods can be applied to study the model using loop quantum gravity. The heat capacity of the universe $C_u$ derived from loop quantum gravity must be less than zero which means the main source of entropy of the universe is gravity. The results for free energy $F_u$ and pressure $p$ are more sutble. Free energy $F_u$ has two stationary points. However, the real minimum cannot be determined in this model since the result depends on the values of $\Lambda (\gamma)$ and $q$ which cannot be derived by this theory. The two stationary points correspond to two different pressures $p$. The results for a Schwarzschild black hole on the other hand are much clearer. In ref.[9], the authors discussed the thermodynamic stability of a black hole using loop quantum gravity.  From this paper we know that in loop quantum gravity, the thermodynamic stability of a Schwarzschild black hole can be determined depending on whether the value of $(1-q)$ is larger than zero or not. We do not list the details. \\
In the future, we intend to  proceed along at least three different lines of further research. Firstly, we can directly generalize our results from the Tsallis $q$-entropy discussed  in this paper to Rényi $q$-entropy. Secondly, the thermodynamic stability of the universe can be  thoroughly studied using fundamental theories, such as Double Field Theory (DFT). Thirdly, since the spherically symmetric solution of a black hole in DFT has been obtained [22],  the thermodynamic properties of this new solution for a black hole can be studied and compared to the results of the Schwarzschild solution.



\begin{thebibliography}{99}

\bibitem{1}
J.D. Bekenstein, \emph{Phys. Rev. D}, {\bf 7} (1973) 2333  

\bibitem{2}
J.D. Bekenstein, \emph{Phys. Rev. D}, {\bf 9} (1974) 3292

\bibitem{3}
S.W. Hawking, \emph{Nature}, {\bf 248} (1974) 30

\bibitem{4}
S.W. Hawking, \emph{Phys. Rev. D}, {\bf 13} (1976) 191

\bibitem{5}
Tamas S. Biro, Viktor G. Czinnera, \emph{Phys. Lett. B}, {\bf 726} (2013) 861-865.

\bibitem{6}
Constantino Tsallis, Leonardo J.L. Cirto, \emph{Eur. Phys. J. C}, {\bf 2487} (2013).

\bibitem{7}
Abhishek Majhi, \emph{Phys. Lett. B}, {\bf 775} (2017) 32-36

\bibitem{8}
Viktor G.Czinner, Hideo Iguchi, \emph{Phys. Lett. B}, {\bf 752} (2016) 306-310

\bibitem{9}
K.Mejrhit, S-E.Ennadifi, \emph{Phys. Lett. B}, {\bf 794} (2019) 45-49

\bibitem{10}
Viktor G.Czinner, Hideo Iguchi, \emph{Eur. Phys. J. C}, {\bf 77} (2017) 892

\bibitem{11}
Ahmad Sheykhi, \emph{Phys. Lett. B}, {\bf 785} (2018) 118-126

\bibitem{12}
C. Tsallis, \emph{J. Stat. Phys}, {\bf 479} (1988) 52

\bibitem{13}
M.L. Lyra, C. Tsallis,  \emph{Phys. Rev. Lett.}, {\bf 80} (1998) 53

\bibitem{14}
C. Tsallis, R.S. Mendes, A.R. Plastino, \emph{Physica A }, {\bf 261} (1998) 534

\bibitem{15}
G. Wilk, Z. Wlodarczyk, \emph{Phys. Rev. Lett.}, {\bf 84} (2000) 2770

\bibitem{16}
A. Saguia, M.S. Sarandy, \emph{Phys. Lett. A}, {\bf 374} (2010) 3384

\bibitem{17}
M. Akbar, R.G. Cai, \emph{Phys. Rev. D}, {\bf 75} (2007) 084003

\bibitem{18}
R.G. Cai, S.P. Kim,  \emph{J. High Energy Phys.}, {\bf 0502} (2005) 050

\bibitem{19}
G. Izquierdo, D. Pavon,  \emph{Phys. Lett. B}, {\bf 633} (2006) 420

\bibitem{20}
Edward W.Kolb and Michael S.Turner, The Early Universe, CRC Press.

\bibitem{21}
Constantino Tsallis, Fulvio Baldovin, Roberto Cerbino and Paolo Pierobon, arXiv:cond-mat/0309093v1

\bibitem{22}
Stephen Angus, Kyoungho Cho, Jeong-Hyuck Park, \emph{Eur. Phys. J. C}, {\bf 78} (2018) 500




\end{thebibliography}
\end{document}